\documentclass[a4paper]{jpconf}
\usepackage{graphicx}
\usepackage{amsmath}
\usepackage{amssymb}

\def\Nbub{\langle {\cal N}\rangle}

\begin{document}
\title{On the cosmological implications of the electroweak vacuum instability: constraining the non-minimal coupling with inflation}

\author{Andreas Mantziris}

\address{Department of Physics, Imperial College London, 180 Queen's Gate, London, SW7 2AZ, UK}

\ead{a.mantziris18@imperial.ac.uk}

\begin{abstract}
Our current measurements of the Standard Model parameters imply that the Higgs field resides in a metastable electroweak vacuum, where the vacuum can decay to a lower ground state, with cataclysmic repercussions for our Universe. According to our observations, no such event has happened in the observable universe, in spite of the various energetic processes that could have triggered it. This work serves as an overview of a method that uses the metastability of the false vacuum during cosmological inflation to provide constraints on the Higgs curvature coupling $\xi$. Considering also the effects of the time-dependent Hubble rate on the effective Higgs potential and on our past light-cone space-time geometry, results in state-of-the-art lower $\xi$-bounds from quadratic and quartic chaotic inflation, and Starobinsky-like power-law inflation.
\end{abstract}

\section{Introduction}

\begin{figure}[h]
\centering
\includegraphics[width=1.\linewidth]{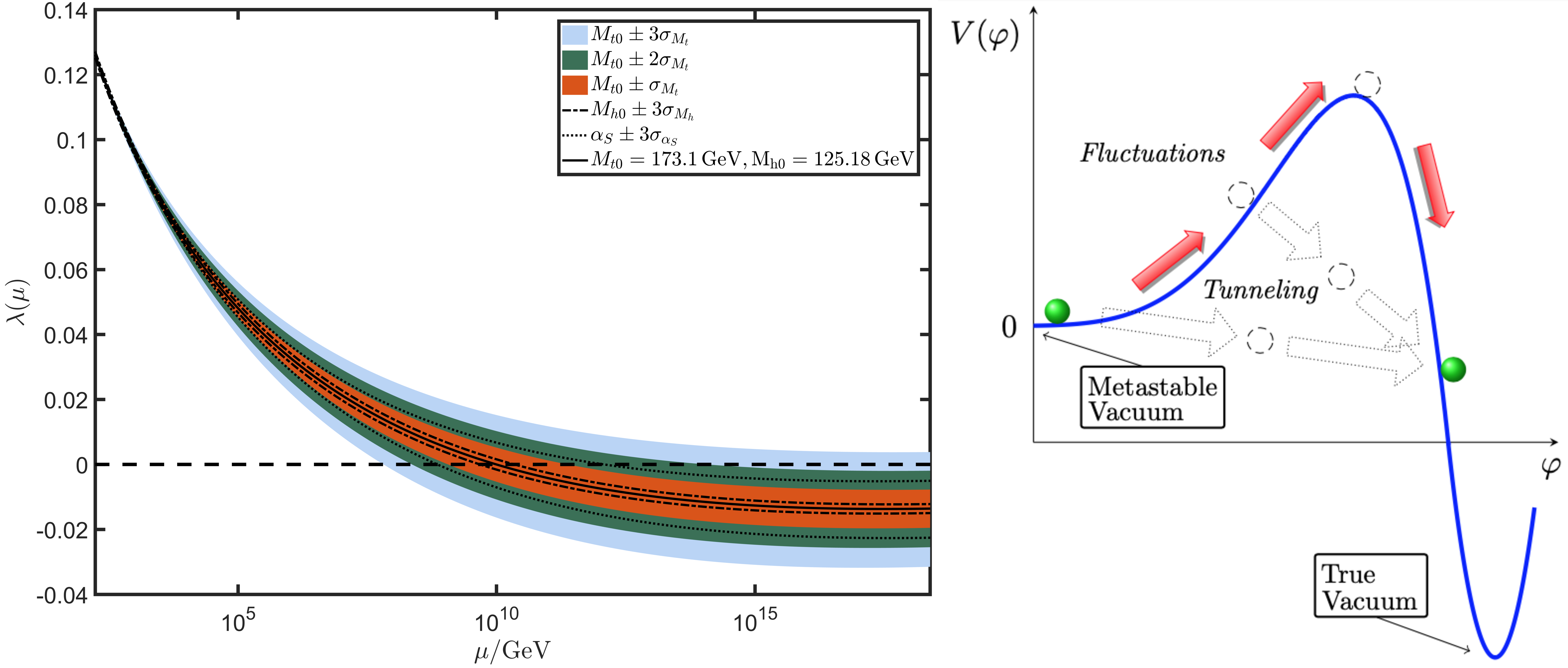}
\caption{Left: running of the Higgs self-coupling, where the coloured bands correspond to the uncertainties in the top quark mass. Right: double well potential of a scalar field $\varphi$, with possible vacuum decay trajectories from a metastable vacuum to the true vacuum state. \cite{Markkanen:2018pdo} }
\label{fig:1}
\end{figure}

The experimental values for the parameters of the Standard Model (SM) of particle physics, especially $m_h$ and $m_t$ (mass of the Higgs boson and the top quark), have some important implications for our Universe. First of all, the Higgs self-interaction $\lambda$ does not diverge below the Planck scale \cite{Bednyakov:2015sca}, as it can be seen in the left plot of figure \ref{fig:1}. It is evident that the four-point coupling runs smoothly with the energy scale $\mu$, throughout the coloured band that signifies the uncertainty in the measurement of the top quark mass. This suggests that the SM can persist as a valid theory of fundamental physics at higher energies and thus provide a consistent minimal model of the early Universe. Secondly, for most of the values within the $3 \sigma$ deviation band, $\lambda$ turns negative during its running. This leads to the formation of a second, deeper vacuum state at higher values of the Higgs field $h$, since the Higgs quartic potential is given by
\begin{align}
	V_H (h, \mu) =  \frac{\lambda(\mu)}{4} h^4  \, ,
\end{align}
with is a potential barrier separating the two vacua, as shown in the right plot of figure \ref{fig:1}. The presence of a lower ground state renders the EW state, that our Universe resides in, to be a metastable one because the Higgs field is prone to vacuum decay. This process can take place via quantum tunnelling, thermal fluctuations, or a combination of both and because our measurements dictate that we are still in the EW vacuum, it means that the metastable state has survived throughout our cosmological history. This realisation inspired us to study vacuum decay in early universe scenarios to constrain fundamental physics. More specifically, we derived lower bounds for the Higgs-curvature coupling $\xi$ from vacuum stability during inflation \cite{Mantziris:2020rzh}.

\section{Vacuum decay during inflation}
Vacuum decay induces the nucleation of true vacuum bubbles expanding with velocity close to the speed of light that consume everything in their way. Since our Universe is still in the EW vacuum, no bubbles could have formed in our observable universe. This can be expressed in terms of the expectation value of the number of bubbles in our past light-cone as $  \langle {\cal N} \rangle \leq 1 $, where
\begin{align}
d \langle {\cal N} \rangle = \Gamma d {\cal V} \, \Rightarrow \, \, \langle {\cal N} \rangle = \int_{\substack{ \rm past }} d^4x \sqrt{-g}\Gamma(x) \,,
\label{eq:Nbubs}
\end{align}
with $\Gamma$ being the decay rate, ${\cal V}$ the space-time volume and $g$ the determinant of the metric.

For our purposes, it is more convenient and intuitive to integrate (\ref{eq:Nbubs}) backwards in time over the duration of inflation, in terms of  $e$-foldings of inflation $N = \mathrm{ln} \left(a_{\rm inf} / a(\eta) \right)$, i.e.
\begin{align}
 \langle {\cal N} \rangle =  \frac{4\pi}{3}\int_0^{N_{\mathrm{start}}} dN \left( \frac{a_{\mathrm{inf}} \left(\eta_0-\eta\left(N\right)\right)}{e^{N}} \right)^3  \frac{\Gamma(N)}{H(N) } \, ,
 \label{eq:Nbub-inf}
\end{align}
where $a_{\rm inf}$ is the scale factor at $N_{\rm inf}=0$ (end of inflation), $\eta_0$ the conformal time today and $H$ the Hubble rate. These quantities are calculated subject to an inflationary potential $V(\phi)$ via
\begin{align}
\frac{d \Tilde{\eta}}{dN} &= - \Tilde{\eta}(N) - \frac{1}{a_{\mathrm{inf}} H(N)}  \, ,
\label{eq:deta-dN} \\
H^2 &=  \frac{V(\phi)}{3M_P^2} \left[ 1 -  \frac{1}{6 M_P^2} \left( \frac{d \phi}{d N} \right)^2 \right]^{-1} \, ,
\label{eq:Hubble-N}
\end{align}
without slow roll approximations, where $\Tilde{\eta}=e^{-N} \eta$ and the inflaton field evolves according to
\begin{align}
\frac{d^2\phi}{dN^2} &=\frac{V(\phi)}{M_P^2H^2} \left(\frac{d\phi}{dN}-M_P^2\frac{V'(\phi)}{V(\phi)}\right) \, .
\label{eq:dphi-dN}
\end{align}
Observations dictate that inflation lasts for at least 60 $e$-folds, i.e. $N_{ \rm start} \geq 60$, and by implementing $  \langle {\cal N} \rangle \leq 1 $, we can constrain the remaining free parameter in (\ref{eq:Nbub-inf}). This is the non-minimal coupling $\xi$, which enters the calculation via the decay rate $\Gamma$ as explained below.

The classical solutions to the decay process from false to true vacuum are famously known as instantons \cite{Callan:1970ze, Coleman:1980aw}. During inflation, the Hubble rate is at such high scales that the Hawking-Moss (HM) instanton \cite{Hawking:1981fz} is the dominant one, characterised by the decay rate $\Gamma_{\rm HM}$ given by
        \begin{align}
             \Gamma_{\rm HM}(R) \approx \left(\frac{R}{12}\right)^2 e^{- B_{\rm HM}(R)}  \, , \quad B_{\rm HM}(R) \approx \frac{384 \pi^2  \Delta V_{\rm H} }{R^2} \, ,
            \label{eq:Gamma}
        \end{align}     
where $R$ is the Ricci scalar, $B$ the action difference and $\Delta V_{\rm H} = V_{\rm H}(h_{\rm bar}) - V_{\rm H}(h_{\rm fv})$ the height of the potential barrier, measured from the top of the barrier (bar) to the false vacuum (fv). We calculate  $\Delta V_{\rm H}$ using the Renormalisation Group Improved (RGI) effective Higgs potential
\begin{align}
    V_{\rm H}^{\rm RGI}(h, R) = \frac{\xi(\mu_*(h, R))}{2}  R h ^2 +\frac{\lambda(\mu_*(h, R))}{4}  h^4 + \frac{\alpha(\mu_*(h,R))}{144} R^2 \, ,
    \label{eq:VRGI}
\end{align}
where the first is the Higgs-curvature term], the second the self interaction and the third a radiatively generated gravitational term (in de Sitter (dS) space) originating from curved space proper renormalisation. The RG scale $\mu_*$ is chosen so that the loop correction to (\ref{eq:VRGI}) vanishes.

\section{Results}

\begin{figure}[h]
\centering
\includegraphics[width=0.6\linewidth]{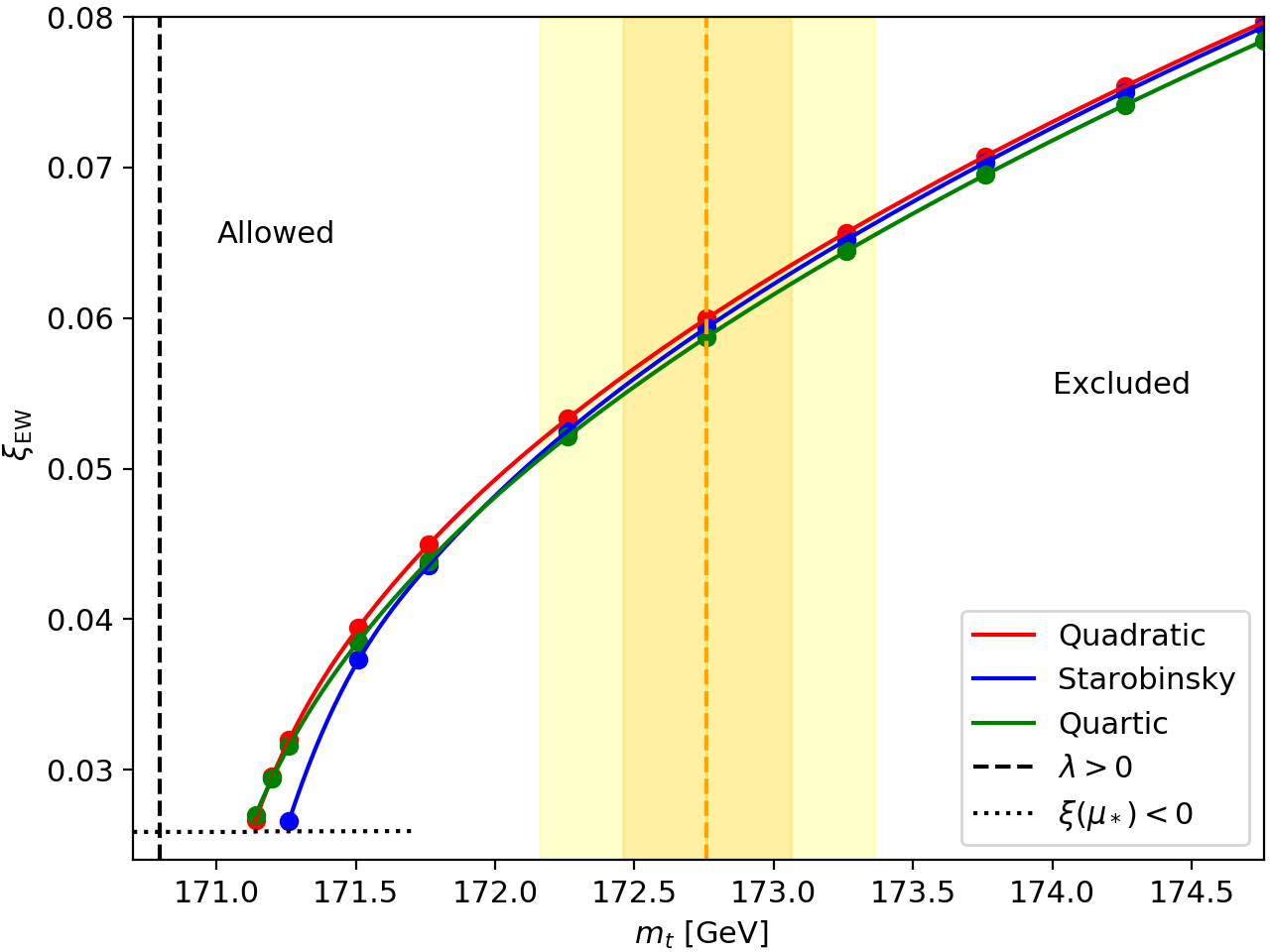}
\caption{Non-minimal coupling bounds with a varying top quark mass for three models. \cite{Mantziris:2020rzh}}
\label{fig:2}
\end{figure}

Solving the system of (\ref{eq:Nbub-inf}), (\ref{eq:deta-dN}) and (\ref{eq:dphi-dN}) for quadratic inflation $V(\phi) = \frac{1}{2}m^2\phi^2$, quartic inflation $V(\phi) = \frac{1}{4} \lambda \phi^4 $, and Starobinsky-like \cite{Starobinsky:1979ty} power-law inflation $V(\phi) = \frac{3}{4} \alpha^2 M_P^4 \left( 1 - e^{-\sqrt{\frac{2}{3}}\frac{\phi}{M_P}} \right)^2$ results in the bounds shown in figure \ref{fig:2}. As $\xi$ is a running parameter, the bounds refer to its value at $\mu_{\rm EW}$. In the lower left corner of the plot, the parameter space is restricted according to the signs of $\lambda$ and $\xi$. If $\lambda$ remains positive during its running, we have stability of the EW vacuum, and therefore we cannot obtain any constraints from metastability. The curves terminate at the horizontal dotted line, because our approach does not take into account any potential destabilization of the Higgs potential caused by a negative $\xi$ during its RG evolution.

\begin{figure}[h]
\centering
\includegraphics[width=0.6\linewidth]{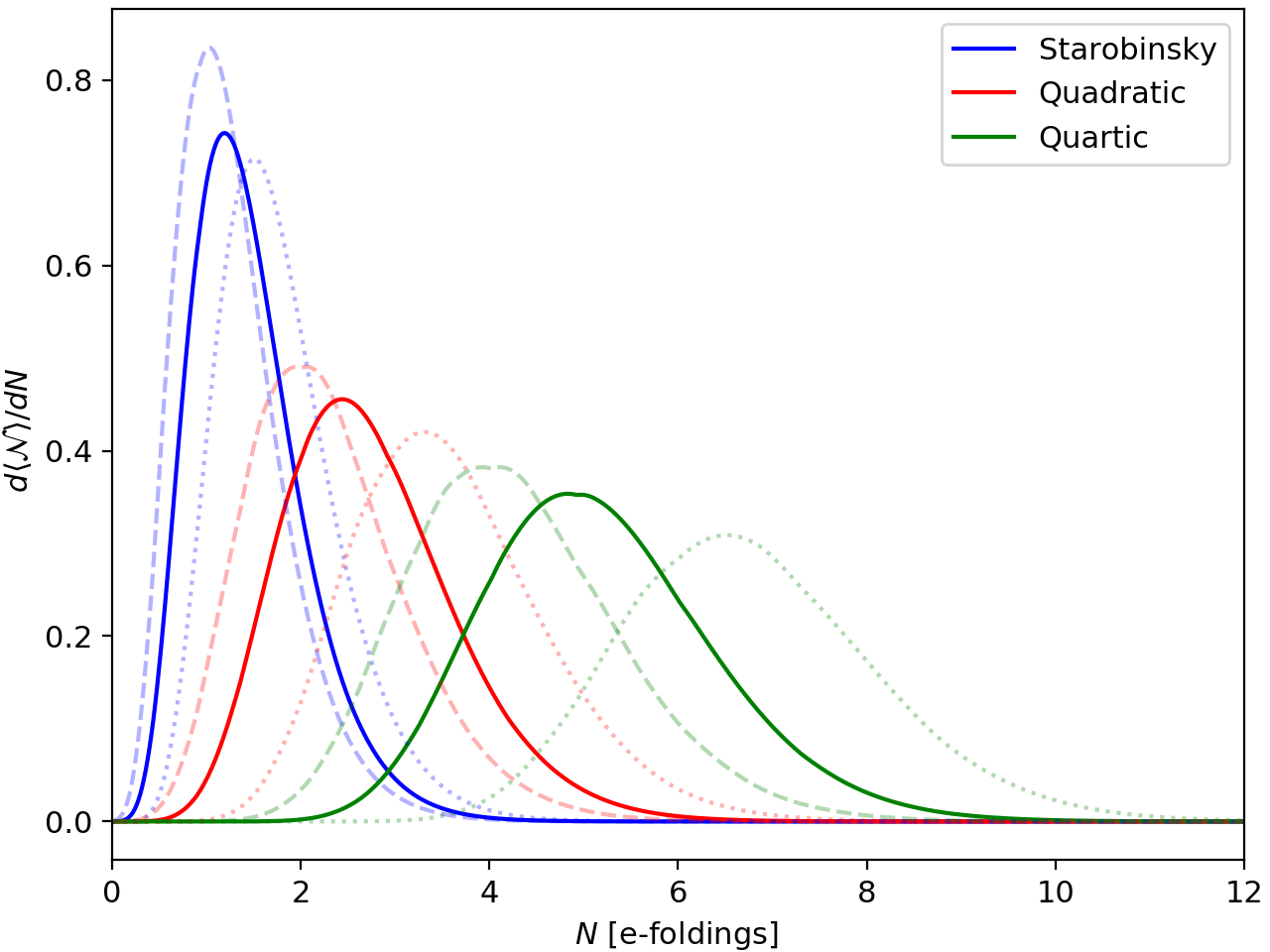}
\caption{Bubble integrands for different inflationary models and top quark masses (central value: solid, $\pm 0.5$ GeV: dashed/dotted), with $\xi_{\rm EW}$ tuned to $\Nbub = 1$ for each case. \cite{Mantziris:2020rzh}}
\label{fig:3}
\end{figure}

In figure \ref{fig:3}, the localised peaks of the integrands of (\ref{eq:Nbub-inf}) for each inflationary model highlight the time when bubble production is most likely to take place. In all cases, the bubbles have nucleated close to the end of inflation but before the last $e$-folding, while a 0.5 GeV variance around the central $m_t$ value is also accounted for and illustrated with the corresponding fainter dashed and dotted lines. It is important to underline the significance of this, because a certain amount of dS approximation has been used in this computation (see \cite{Mantziris:2020rzh} for details) and space-time does not resemble approximately dS towards the very end of the inflationary expansion. Moreover the unknown total duration of inflation does not affect these results, as bubble formation is suppressed unless inflation lasts for an unrealistic duration of $\sim 10^{60}$ $e$-folds. Thus, these bounds are trustworthy, within the range of validity of our approximations and assumptions.

\section{Conclusions}

With the method presented here, we accounted for 1-loop curvature corrections beyond dS in the RGI effective Higgs potential, to result into model degenerate, duration independent and $m_t$-dependent lower bounds on the Higgs-curvature coupling, which are summarised as $\xi_{\rm EW} \gtrsim 0.06$. Furthermore, the dependence of predominant bubble production on the total duration of inflation hints against eternal inflation, but further study is required to address this point properly.

\ack
This work was conducted under the supervision of Arttu Rajantie and in collaboration with Tommi Markkanen, with the author being supported by an STFC PhD studentship.

\section*{References}
\bibliography{references.bib}

\end{document}